\documentclass{kapproc} 
\usepackage{psfig}

\kluwerbib

\begin{document}



\articletitle{Cosmology with Galaxy Clusters through the Sunyaev-Zel'dovich 
effect}


\author{J.M Diego}
\affil{Physics Dept. Denys Wilkinson Bldg, Keble Road, Oxford OX1 3RH, U.K.}
\email{jmdr@astro.ox.ac.uk}

\begin{abstract}
Future Sunyaev-Zel'dovich effect (SZE) surveys will detect 
thousands of clusters, many of them at high redshift.  
We show how the SZE number counts as a function of flux can be used 
as a useful way of constraining the cosmological model. 
The constraints can be dramatically improved if 
redshift information is provided for a small subsample of the observed 
clusters. In this work we present the constraints on $\Omega_m$ and 
$\sigma_8$ expected for a combined mm-optical survey. 
The mm survey (Planck) providing the cluster number counts as a function of 
flux and the optical survey (GTC) the redshift estimation of a small subsample 
of the cluster catalogue. We also show how it is possible to get an 
accurate estimate of the cluster flux using a non-parametric component 
separation algorithm and how using morphological redshifts it is possible 
to make a high-z preselection of the small subsample to be observed in 
the optical.
\end{abstract}


\section{Constraining the cosmology with the SZE.}
Studies on the local abundance of clusters have shown that 
today's observed cluster density constrains the cosmological 
model. However, these constrains show a degeneracy between the matter 
density, $\Omega_m$, and the normalisation of the power spectrum, 
$\sigma_8$ (Bahcall \& Cen 1993). 
This degeneracy can be broken by including in the analysis 
the cluster abundance at high redshift. However, observations of 
clusters at high redshift are difficult because of the decrease of  
flux with redshift (flux $\propto D_l(z)^{-2}$)
The peculiarity that the central decrement of the SZE is redshift 
independent, makes this effect an interesting way of observing high 
redshift clusters. 
Future mm experiments, like Planck, will detect thousands of clusters. 
A large proportion of these clusters will be at high redshift. Given 
such a large number of clusters, it is natural to think that the 
cluster catalogue provided by these experiments will be an important 
source of cosmological studies. However, since the SZE is redshift 
independent, these catalogues will contain no information about the 
redshift of the clusters. They will contain information only about 
the flux of the clusters. With the flux, it is possible to build the 
cluster number counts as a function of flux. 
By using this data set alone, one can not, for instance, distinguish 
between a $\Lambda$CDM and an OCDM models both with the same matter 
content. Redshift information is essential to distinguish between these 
two models. However, we do not need to know the redshift of all the 
clusters in order to distinguish between these two models. An optically    
identified subsample of about 300 clusters (in the case of Planck) 
{\it randomly} selected from the SZE catalogue is enough to distinguish 
between the above models at a $3\sigma$ level (see details in 
Diego et al. 2002a). By {\it randomly} we mean that from the SZE catalogue we 
pick up 300 clusters randomly selected from the full catalogue. In this 
random subsample there will be many low, intermediate and some high redshift 
clusters.
\begin{figure}[ht]
\centerline{\psfig{file=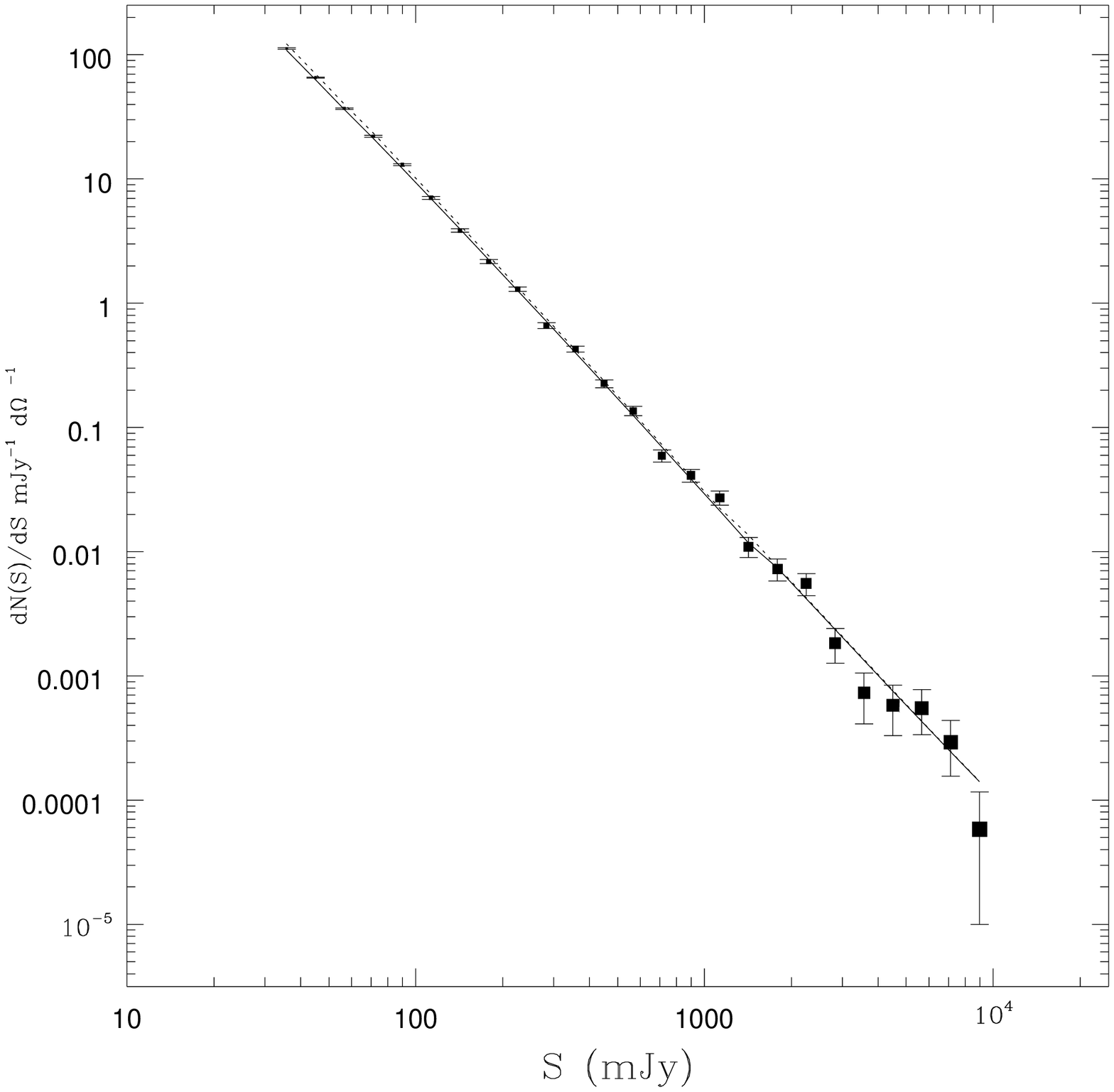,width=2.5in} 
            \psfig{file=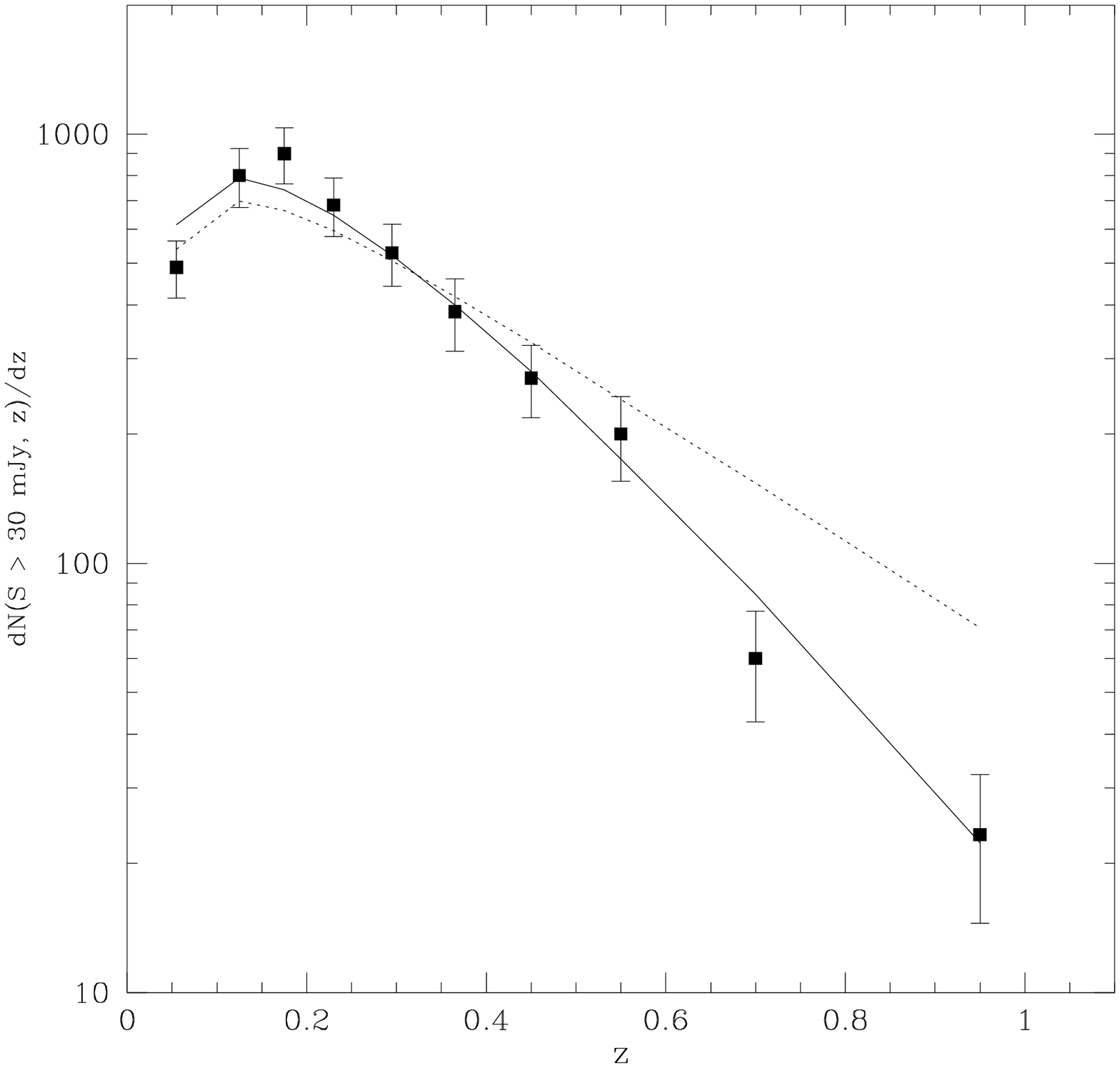,width=2.5in}}
\caption{Taken from Diego et al. 2002a. 
Predicted cluster number counts for Planck (left) and number 
counts per redshift bin (right) for the GTC followup of the small 
subsample of 300 clusters. 
The lines show the predictions for the theoretical model. 
$\Lambda$CDM solid line and OCDM dotted line ($\Omega_m = 0.3$, 
$\sigma_8 = 0.8$ in both cases).}
\end{figure}
We illustrate this situation in figure 1 where we compare the above models 
with the expectations for the Planck satellite (left) and the optical 
followup of the subsample with 300 clusters performed by a 10-m class 
telescope like GTC. 
Using the data sets of figure 1 it is possible to constrain $\sigma_8$ and 
$\Omega_m$ within an accuracy of about 0.1 (Diego et al. 2002a). 
However, to get the data sets of figure 1 we need first to detect the clusters 
and estimate their flux and second to estimate the redshift of a small 
subsample ($\approx$ 300 clusters) of clusters. In the next two sections we 
present two techniques which will be useful to solve this problems.
\section{Cluster detection: Non-parametric method}
In this section we will give a brief description of the non-parametric method 
presented in Diego et al. 2002b which has demonstrated its robustness with 
simulated Planck data. 
The SZE produces a distortion in the CMB spectrum shifting the CMB photons 
to higher frequencies. This shift can be {\it easily} observed if we have 
multifrequency maps of the CMB. However multifrequency maps of the CMB also 
contain other components: Galactic, extragalactic and instrumental noise.  
The separation of all these components is the main aim of the different proposed 
component separation algorithms. Due to the large number of components involved in 
this problem, the component separation algorithms usually need to make many 
assumptions about the frequency dependence of each component and it  
angular power spectrum. These assumptions can produce a bias in the final 
result if some of the assumptions are wrong. There is still a lot of uncertainty 
about the frequency dependence of several components and/or about their angular power 
spectrum. The non-parametric method proposed in Diego et al. 2002b overcomes 
these problems by doing no assumptions except for the frequency dependence 
of the SZE (which is well known) and a weak assumption about the prior (however, 
the results seem to be almost insensitive to this assumption). 
In a few words, the non-parametric method deconvolves an optimally weighted 
combination of the multifrequency maps which have been previously 
{\it pre-processed} in order to increase the SNR of the SZE. When applied to 
Planck simulated data, the non-parametric method predicts that about 10000 
clusters will be detected in all the sky. We show in figure 2 (left) how the 
non-parametric method is able to recover the fluxes which are our main interest 
to build the first of the data sets in figure 1. 
\section{Optical followup: Morphological redshifts}
Although in figure 1 we assumed that the subsample of 300 clusters 
was randomly selected from the SZE catalogue, the results would improve if 
the ratio of high-to-low redshift clusters is high. In order to achieve this 
high ratio it would be interesting to preselect the small subsample by 
redshift instead of performing a random preselection. But how to make this 
redshift preselection if the SZE is redshift independent ? 
In fact, although the spectrum of the SZE is redshift independent, there are 
observables like the total flux which do depend on the redshift of the cluster. 
By combining different SZE observables it is possible to get an estimate 
of the redshift (Diego et al. 2002c). 
Morphological redshifts can provide this estimate combining several 
observables in a principal component analysis (PCA). The principal 
components (pc) are then calibrated with a set of known redshifts and then a
Bayesian estimator is used to get the most likely redshift of the clusters. 
In figure 2 (right) we show the performance of this technique applied using 
N-body high resolution simulations. These redshift estimates can then be used to 
preselect a subsample of high redshift cluster candidates or to optimise 
the followup of a randomly selected subsample by observing the low redshift 
candidates with small telescopes and the high redshift clusters with large 
telescopes. 
\begin{figure}[ht]
\centerline{\psfig{file=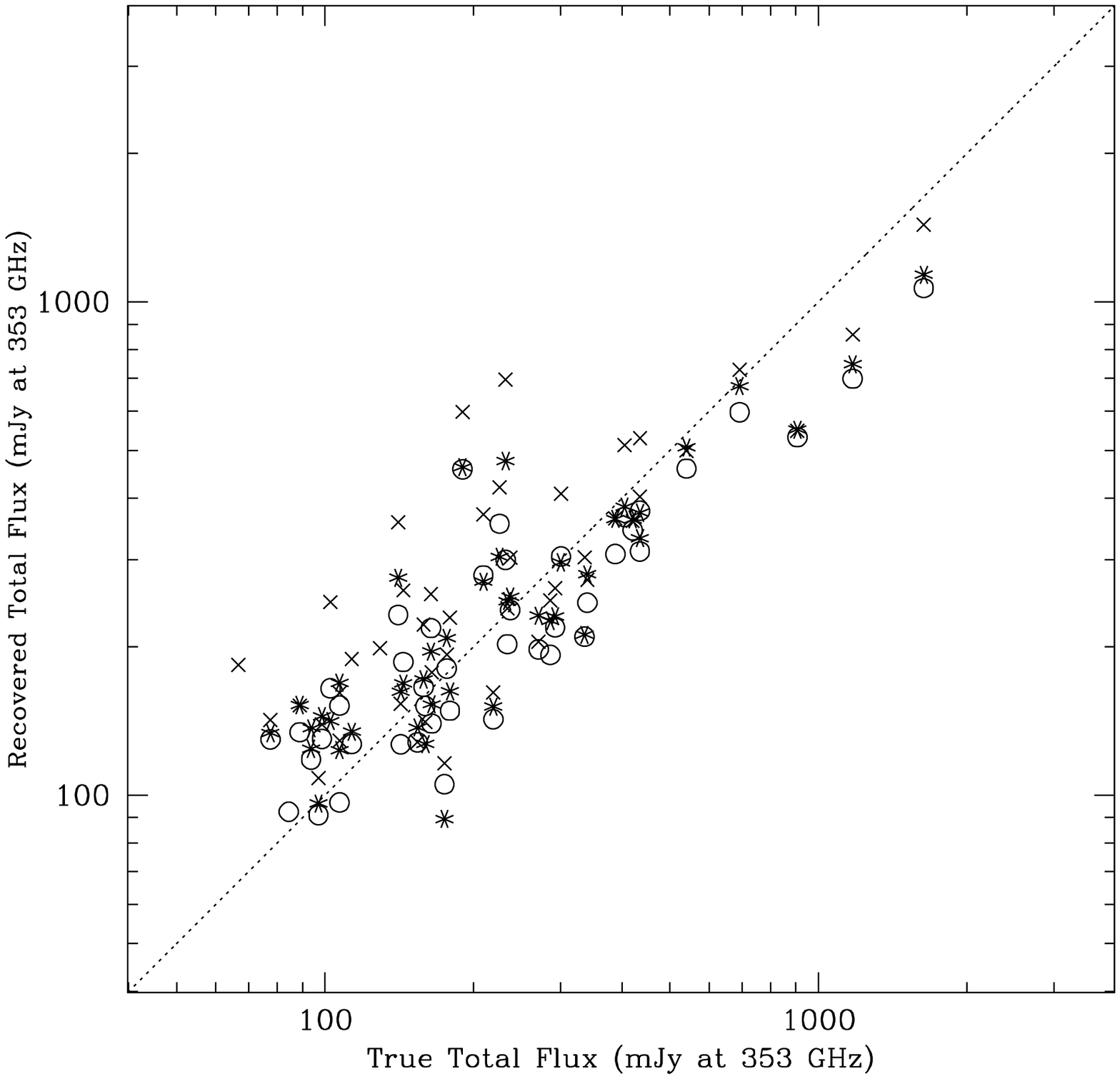,width=2.5in} 
            \psfig{file=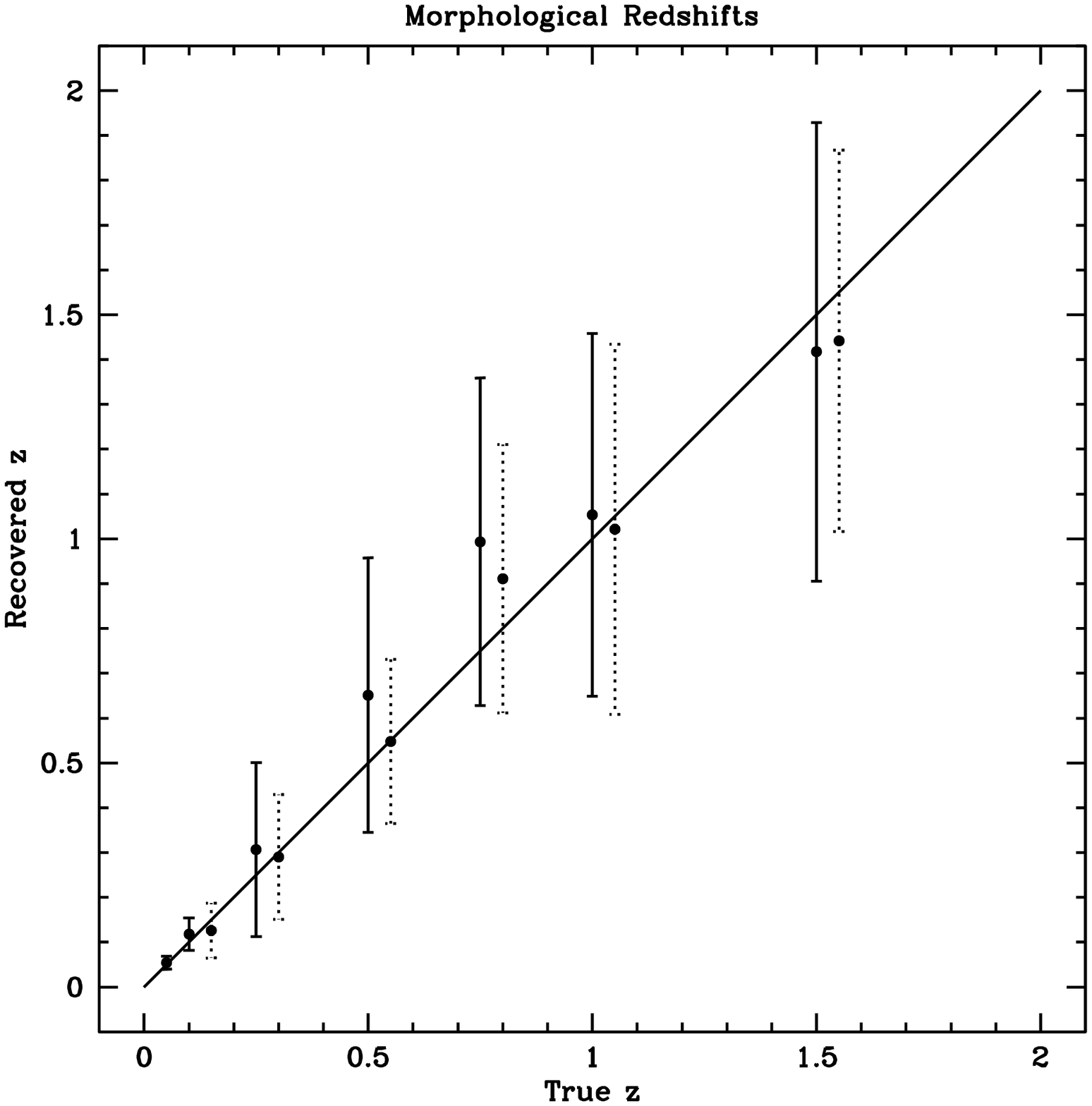,width=2.5in}}
\caption{{\bf Left}. Taken from Diego et al. 2002b. 
Stability on the recovered fluxes. The different symbols correspond 
to different assumptions on the prior. The result shows to be quite insensitive 
to this assumption demonstrating the robustness of the method.
{\bf Right}. Taken from Diego et al. 2002c. 
Performance of the morphological redshifts for a survey with 
FWHM = 25 arcsec resolution. The method improves if only the first pc is used 
in the high redshift interval (dotted errors). However, at low redshifts, the other pcs help 
to reduce the error in the $z$-estimation (solid errors).}
\end{figure}


\end{document}